 \documentstyle[a4,12pt]{article}
\renewcommand{\baselinestretch}{1.2}
\def\scf{superconformal}
\def\scfty{superconformality}
\def\sd{superdiffeomorphism}
\def\sc{supersymmetric}
\def\sdf{superdifferential}
\def\sdfs{superdifferentials}
\def\sz{supersymmetrization}
\title{{\bf The Polyakov action on the supertorus}}
\author{{\bf Jean-Pierre Ader and Hamid Kachkachi} \\
{\em Laboratoire de Physique Th\'{e}orique,$^{\dag}$} \\
{\em Universit\'{e} de Bordeaux I, }\\
{\em 19 rue du Solarium, F-33175 Gradignan Cedex}}
\begin{document}
\begin{titlepage}
\maketitle
\thispagestyle{empty}
\vspace{1 cm}
\begin{abstract}
A consistent method for obtaining a well-defined Polyakov action on the
supertorus is presented.
This method uses the covariantization of derivative operators and
enables us to construct a Polyakov action which is globally defined.
\end{abstract}
\vspace{4cm}
LPTB 92-7\\
August 1992
BITNET Addresses: {\bf Ader@FRCPN11}, {\bf Kachkach@FRCPN11}\\[2cm]
$\dag$ $\overline{\mbox{Unit Associe au CNRS}}$, U.A.764
\end{titlepage}
\section{Introduction}
Recently the general expressions of the chirally split superdiffeomorphism
anomaly and the associated effective action on the $N=1$ superplane (SC)
have been found \cite{DelGie}. This action, a generalization of
the Polyakov action, obeys the anomalous Ward identity

\begin{equation} \label{1}
 s\Gamma_{WZP}(H,\bar{H}) =  \int d^{4}z {\cal A}(C^{z},H) + c.c ,
\end{equation}
where $s$ is the BRST operator and ${\cal A}(C^{z},H) + c.c$ is
the chirally split form of the superdiffeomorphism anomaly. In this formula
appear the super Beltrami coefficient $H$ (and its complex conjugate
$\bar H$) and the ghost fields $C^{z}$,
$C^{\bar{z}}$ which parametrize superdiffeomorphism transformations.
The two terms in the r.h.s. separately
satisfy the Wess-Zumino (WZ) consistency condition.
In eq.(\ref{1}) a special gauge choice for the super Beltrami coefficients
is made which has been proved to be always possible \cite{Tak}. This choice
reduces the number of independent Beltrami coefficients to two
( $H$ and its c.c partner ).\\
The characteristic feature of the superdiffeomorphism anomaly, the
factorized form in eq.(\ref{1}), results in the holomorphic factorization
property of the functional $\Gamma_{WZP}$ which reads
\begin{equation}\label{2}
\Gamma_{WZP}(H,\bar{H}) =  \Gamma_{WZP}(H) +  \overline{\Gamma_{WZP}(H)}.
\end{equation}

      Extension to a generic super Riemann surface has been
done only for the superdiffeomorphism anomaly \cite{DelGie}. In contrast, the
corresponding problem in the bosonic case has been solved on the torus
by Lazzarini \cite{Laz} and the generalization to a Riemann
surface of any genus has been recently given by Zucchini \cite{Zuc}.
The purpose of this work is to derive an expression for the functional
$\Gamma_{WZP}$ which is well-defined on the supertorus. Thus, it can be
viewed as the supersymmetric extension of the result of ref.\cite{Laz}.\\
      First, in sect.2, we introduce
some basic notions on super Riemann surfaces
(SRS) and superdifferentials necessary for our purpose; most important is
the notion of covariant derivative associated to an affine connection,
since it allows us to construct actions on a
generic Riemann surface in a systematic and straightforward way. Thus
it seems natural, at first, to present the result of ref.\cite{Laz}
; although the derivation in this case seems particularly simple,
it will serve as an illustrative example of the efficiency of the method,
sect.3. The extension to the supersymmetric case is given in sect.4 for
the supertorus. The component content of our formulation is presented in
sect.5 with a special emphasis on the parametrization of the isothermal
super-coordinates. Finally we end with some remarks.

\section{Mathematical preliminaries}
Before proceeding to the main body of this paper we shall give a brief
overview of the already established features of a SRS necessary for our
purpose. We will also give examples of some objects related to these features
to fix the notation.

\subsection{Superdifferentials}
Let us first define the framework in which we are working.
We consider a supermanifold $\hat{M}$ which is obtained by patching together
local coordinate charts $(U,(z,\theta))$, where $U$ is an open subset
of the underlying manifold $M$ of $\hat{M}$ obtained by switching off
the nilpotent elements of $\hat{M}$;
 $(z,\theta)$ is a pair of complex coordinates with $\theta$ anticommuting.
(see \cite{Rog,RabinCrane}).
The complex supermanifold thus defined becomes a SRS if the transition
functions between two local coordinate charts $(U,(z,\theta))$ and
$(V,(\tilde{z},\tilde{\theta}))$
satisfy the following conditions of \scfty

\begin{equation}\label{3}
\left\{\begin{array}{lll}
 \tilde{z}&=&\tilde{z}(z,\theta) \\
 \tilde{\theta}&=&\tilde{\theta}(z,\theta) \\
 D_{\theta}\tilde{z}&=&\tilde{\theta}D_{\theta}\tilde{\theta}.
 \end{array}\right.
 \end{equation}

 where $D_{\theta}=\partial_{\theta}+\theta \partial_{z}$ is the
 superderivative
 obeying $(D_{\theta})^2=\partial_{z}\equiv{\partial}$. It is understood that
 the complex conjugate (c.c) conditions are also to be taken into account.

 An atlas of superprojective coordinates
 $(\hat{Z},\hat{\Theta},\hat{\bar{Z}},\hat{\bar{\Theta}})$
 on a SRS (without boundary)  $\hat{\Sigma}$ defines  supercomplex structures,
 or equivalently, superconformal classes of metrics, related to the reference
 structures $(z,\theta,\bar{z},\bar{\theta})$ by the super Beltrami
 differentials
 via the super Beltrami equations, refs.\cite{Tak,RabinCrane,DelGie'}.
 In general these structures
 are parametrized by two independent odd superfields
 $H^{z}_{\bar{\theta}}$ and $H^{z}_{\theta}$ ( and the c.c analogs).
The super Beltrami field $H^{z}_{\bar{\theta}}\equiv H$ contains the
 ordinary Beltrami
 coefficient $\mu$ and its fermionic partner the beltramino $\alpha$, whereas
 $H^{z}_{\theta}$ contains only auxiliary space-time fields. This is the
 reason why, in general,
 studies are limited to the special case $H^{z}_{\theta}=0$, a restriction
 that we adopt henceforth, and which is equivalent to the following condition

 \begin{equation} \label{4}
   D_{\theta}\hat{Z}\;=\;\hat{\Theta}D_{\theta}\hat{\Theta}.
 \end{equation}

 Now consider the compact SRS $\hat{\Sigma}$ of genus $g$, i.e. with a
 compact underlying
 Riemann surface $\Sigma$ of the same genus. Then let $\hat{\omega}$ be the
 canonical line bundle
 whose generator is $(dz\mid d\theta) $\footnote{From now on, a comma
 will separate holomorphic from antiholomorphic; while a vertical line will
 be used to separate
 even from odd variables.}. This is the dual of the tangent
 bundle $T\hat{\Sigma}$ generated by the vector field $D_{\theta}$ (see
 \cite{Nel,RSV}.)

 The transition functions of $\hat{\omega}$ and $T\hat{\Sigma}$  are
 respectively

 \begin{equation}\label{5}
  F=(D_{\theta}\tilde{\theta})\equiv\exp{(-w)}
 \end{equation}
  and $F^{-1}$,
  in a change of coordinates from the chart $(U,(z,\theta))$ to the chart
 $(V,(\tilde{z},\tilde{\theta}))$.
 This leads to the construction of the globally-defined operator (analog of
 the Dolbeault
 operator of the bosonic theory) $\hat{D}\equiv (dz\mid d\theta)\otimes
 D_{\theta}$ (see ref.\cite{Nel}).

 A $(\frac{p}{2},\frac{q}{2})-$superdifferential $\Phi$ is a field of
 conformal weight
 $(\frac{p}{2},\frac{q}{2})$ on a SRS $\hat{\Sigma}$
 given by a collection of functions (its coefficients)
 ${\phi(z,\theta,\bar{z},\bar{\theta})}$
 (one for each coordinate chart) obeying

 \begin{equation} \label{6}
 \phi(\tilde{z},\tilde{\theta},\tilde{\bar{z}},\tilde{\bar{\theta}})\;=\;
 F^{-p}\bar{F}^{-q}
 \phi(z,\theta,\bar{z},\bar{\theta})
 \end{equation}
 in $(U,(z,\theta))\cap (V,(\tilde{z},\tilde{\theta}))$.

 In the fiber bundle language, $\Phi$ is a section of the cross fiber bundle
 $\hat{\omega}^{\otimes p}\otimes \bar{\hat{\omega}}^{\otimes q}$
 (see \cite{Nel})
 and then is written as
 \begin{equation}
 \Phi(z,\theta,\bar{z},\bar{\theta})\;=\;\phi(z,\theta,\bar{z},\bar{\theta})
 (dz\mid d\theta)^p \otimes (d\bar{z}\mid d\bar{\theta})^q.
 \end{equation}
 Now it is easy to see that $(\frac{1}{2},\frac{1}{2})-$superdifferentials
 can be integrated
 over SRS in the same way that $(1,1)-$differentials are integrated on
 Riemann surfaces, since this yields a well-defined expression.
      Let us now introduce some basic definitions which are essentially
borrowed to F.Gieres \cite{Gie} and recalled here for completness.

 \subsection{Superconnections}

In the sequel, unless otherwise stated, the holomorphy and
the superholomorphy properties are understood w.r.t. the $\mu-$structure
 ($Z$) and the $H-$structure ($\hat{Z},\hat{\Theta}$) respectively.
 Moreover, abelian and superabelian differentials are defined w.r.t. these
structures. The reference structures ($z$) and ($z$,$\theta$) will also be
refered to as the ($\mu = 0$)-structure and ($H = 0$)-structure respectively
, or briefly as the 0-structures.\\
A superaffine connection $\zeta$ is a collection
   $\{ \zeta_{\theta}(z,\theta)\}$
of superholomorphic functions in the $H-$structure $\zeta_{\theta}$, i.e.
$D_{\bar{\Theta}}\zeta_{\Theta}=0$, transforming under a superconformal
change of coordinates in $(U,(z,\theta))\cap (V,(\tilde{z},\tilde{\theta}))$
as follows
\begin{equation}\label{8}
   \zeta_{\tilde{\theta}}(\tilde{z},\tilde{\theta})\;=\;\exp{(w)}
   (\zeta_{\theta}(z,\theta)\;-\;D_{\theta}w).
   \end{equation}

 The $(\frac{1}{2},0)$-superdifferential defined by the coefficient
 \begin{equation}\label{9}
 \eta_{\theta}\equiv D_{\theta}\hat{\Theta},
 \end{equation}

  allows us to build a superaffine connection through the definition
   \begin{eqnarray} \label{10}
   \zeta_{\theta}&=&-\;D_{\theta}\log{(\eta_{\theta})}.
   \end{eqnarray}

  In a superholomorphic reference structure $(\hat{Z}_{0}(z,\theta),
  \hat{\Theta}_{0}(z,\theta))$, the definition (\ref{10}) yields a holomorphic
  superaffine connection $\zeta_{0}$ with, $\bar{D}\zeta_{0}=0$.\\

     Related to the notion of a superaffine connection is that of a
   superprojective connection.
   This is a collection of holomorphic functions
   $\{ R_{z\theta}(z,\theta)\}$
   i.e. $D_{\bar{\Theta}}R_{Z\Theta}=0$, transforming
   under the change of coordinates from $(U,(z,\theta))$ to
   $(V,(\tilde{z},\tilde{\theta}))$ as follows

   \begin{equation}\label{12}
   R_{z\theta}(\tilde{z},\tilde{\theta})\;=\;\exp{(3w)}
   (R_{z\theta}(z,\theta)\;-\; \nonumber
   S(\tilde{z},\tilde{\theta};z,\theta)),
   \end{equation}

   where $S(\tilde{z},\tilde{\theta};z,\theta))$ is the super Schwarzian
   derivative.

   A superprojective connection is obtained from a superaffine connection
   $\zeta$ by
    \begin{equation}\label{RR}
     R_{z\theta}\;=\;-\;\partial_{z}\zeta_{\theta}\;-\;\zeta_{\theta}
     D_{\theta}\zeta_{\theta}.
    \end{equation}

    In analogy to the bosonic case, the difference of two superaffine
    connections
    is a superabelian differential and the difference of two superprojective
    connections
    (or quasi-super\-quadratic differentials) is a superquadratic differential.

    Now we give the differential equation satisfied by
    a $(\frac{j}{2},0)-$ superdifferential $\Xi$.
    In the $H-$structure \cite{Gie}, it reads

    \begin{equation}\label{jed}
    \begin{array}{l}
    \left [ D_{\bar{\theta}}\;-\;H^{z}_{\bar{\theta}}\partial_{z}\;+\;
    \frac{1}{2}(D_{\theta}
     H^{z}_{\bar{\theta}})D_{\theta}\right ]\Xi\;=\;\frac{j}{2}(\partial_{z}
     H^{z}_{\bar{\theta}})\Xi.
    \end{array}
    \end{equation}

Finally the covariant derivative $\nabla$ associated to an
affine connection $\xi$ is defined by \cite{BFIZ,Gie}
   \begin{equation}\label{cd1}
   \nabla\equiv\partial-p\xi,
   \end{equation}

   where $p$ is the conformal weight (relative to the $z-$index) of the
   tensor on which
   $\nabla$ is applied.\\

 Similarly, a covariant superderivative $\hat{\nabla}_{\zeta}$ is associated
 to a superaffine connection $\zeta$ in eq.(\ref{10}) by defining its action
 on supertensors
 of conformal weight $\hat{p}$ corresponding to the $(z,\theta)$ sector
   \footnote{In fact the derivatives
   $\nabla$ and $\hat{\nabla}$ appearing in the above equations are the
   coefficients of the operators $\nabla=\nabla_{z} dz$ and
   $\hat{\nabla}=\hat{\nabla}_{\theta}(dz\mid
   d\theta)$
   respectively. This means that the above equations have been written in
   components without displaying
   indices;e.g. $D$ is to be understood as $D_{\theta}$}
 (for instance, $\hat{p}(H)=-1, \hat{p}(D)=\frac{1}{2})$ by \cite{Gie}

   \begin{equation}\label{csd1}
   \hat{\nabla}_{\zeta}\equiv D+2\hat{p}\zeta,
   \end{equation}
   so
   \begin{equation}\label{csd1'}
   \hat{\nabla}_{\zeta}^{2}=\partial+2\hat{p}D\zeta+\zeta D.
   \end{equation}

\section{Covariantization of the Polyakov action on the torus}
Starting from the WZP action on the complex plane \cite{Laz}

\begin{eqnarray*}
\Gamma[\mu]=-\frac{1}{2}\int_{\bf C} d^{2}z \;\mu\partial^{2}log\partial Z
\end{eqnarray*}
and following this reference, we rewrite it as the sum of two terms after an
integration by parts
\begin{eqnarray*}
\Gamma[\mu]=\frac{1}{2}\int_{\bf C} d^{2}z \;A_{T},
\end{eqnarray*}
with
\begin{equation}\label{13}
A_{T} = -\mu\partial\xi-\frac{1}{2}\xi\partial\mu
\end{equation}
where
\begin{equation}\label{14}
\xi=\partial\log{(\partial Z)}
\end{equation}
is a holomorphic affine connection.\\
Each term in the r.h.s. of eq.(\ref{13}) becomes separately globally
defined on the torus in two steps.
First the second term is covariantized by substituting $\nabla$ for
$\partial$ according to eq.(\ref{cd1}), whereas in the first term
the derivative of the affine connection $\xi$ is replaced by the
projective connection associated to it

\begin{eqnarray*}
\gamma = \partial\xi-\frac{1}{2}\xi^{2} .
\end{eqnarray*}

We thus find
\begin{equation}\label{15}
A_{T}=\mu(-\gamma)\;-\;\frac{1}{2}\xi\nabla\mu.
\end{equation}
 Therefore the first term can be made globally defined by
introducing a generic holomorphic
projective connection $r$ (inert under the BRST operator) knowing
that $\mu$ transforms homogeneously under a change of coordinates.
This is due to the fact that the resulting
term is the difference of a generic projective connection $r$ and a
particular one $\gamma$, that is a quadratic differential.\\
The second term, where the partial derivative of the Beltrami differential
$\mu$ has been covariantized to a globally defined expression:
\begin{eqnarray*}
\nabla\mu=(\partial+\xi)\mu,
\end{eqnarray*}
requires in addition the introduction of a holomorphic affine connection
$\xi_{0}$ in the $0-$structure
and then changing $\xi$ to $\xi-\xi_{0}$
which is an abelian differential
\footnote{$\xi_{0}$ is defined according to eq.(\ref{14}) by using the
reference
structure coordinate $Z_{0}$, satisfying $\bar{\partial}Z_{0}=0$.}.\\
Finally we get the globally defined integrand
\begin{equation}\label{16}
A_{T}=\mu(r-\gamma)\;-\;\frac{1}{2}(\xi-\xi_{0})\nabla\mu,
\end{equation}
or explicitly
\begin{eqnarray*}
A_{T}=\mu(r-\partial\xi+\frac{1}{2}\xi^{2})\;-\;
\frac{1}{2}(\xi-\xi_{0})(\partial+\xi)\mu.
\end{eqnarray*}

This is the final expression for the Polyakov action on the torus
found by Lazzarini \cite{Laz};
the action of the BRST operator on it yields the globally defined
(non-integrated) anomaly, (see \cite{LazStora})

\begin{eqnarray*}
\frac{1}{2}(c\partial^{3}\mu-\mu\partial^{3}c)+r(c\partial\mu-\mu\partial c),
\end{eqnarray*}

where $c$ is the ghost parametrizing ordinary diffeomorphisms.\\
One should note here, as was mentioned in the Introduction, that using the
notion of a covariant derivative allows to straightforwardly guess the
globally defined action on the torus.
Let us now perform the equivalent construction on the supertorus.

\section{The globally defined Polyakov action on the supertorus}

The superspace generalization of Polyakov's chiral gauge action
\cite{Pol} has first been found by Grundberg and Nakayama \cite{GN}.
A very compact expression of this functional given in ref.\cite{DelGie}
reads
\begin{equation}\label{17}
\Gamma[H]=\int_{\bf SC} d^{4}z \partial\zeta H,
\end{equation}
where $\zeta$ (in fact $\zeta_{\theta}$) is the coefficient of
the superaffine
connection defined in eq.(\ref{10}).

In the following we will simplify the notation by dropping all indices,
knowing that we use the coefficients of differentials and connections
instead of the fields themselves.

The BRST variation of this functional yields the chirally split
superdiffeomorphism anomaly,
exhibiting the non-invariance of the Polyakov action under a
subgroup \footnote {This subgroup is defined by the restriction
$H_{\theta}^{z}=0$} of the superdiffeomorphism group $SDiff_{0}(\Sigma)$.
 This anomaly reads

\begin{equation}\label{18}
{\cal A}(C,H)=C\partial^{2}DH+H\partial^{2}DC,
\end{equation}

where $C$ is the superdiffeomorphism ghost field.
It is easy to verify by using the following law

\begin{equation}\label{19}
s\zeta=-\frac{1}{2}D\partial C+C\partial\zeta+\frac{1}{2}(\partial C)
\zeta+
          \frac{1}{2}(DC)D\zeta,
\end{equation}
and the BRST transformation of $H$ given in ref.\cite{Gie}
that indeed the action of the BRST operator on eq.(\ref{17}) yields
eq.(\ref{18}).
The $s$ operator is assumed to act from the right; the BRST algebra is
graded by the ghost number, but does not feel the Grassmann parity.

 In order to extend properly the action (\ref{17}) onto the supertorus,
 we first have to show that the integrand can be transformed to a
 globally defined expression by replacing the derivative of the
 superaffine connection by the superprojective connection itself,
by covariantizing the appropriate derivatives appearing in it,
and then introducing the new connections $R$ and $\zeta_{0}$.
Thus the globally defined (non-integrated) anomaly reads\cite{DelGie}

\begin{equation}\label{20}
{\cal A}(C,H,R)=(C\partial^{2}DH+H\partial^{2}DC)+3R(C\partial H-H\partial
C)+DR(CDH+HDC).
\end{equation}

Next we have to verify that this action indeed solves the Ward identity
(\ref{1}), with the above anomalous term. \\
For the first task we write the action in eq.(\ref{17}) as the sum of two
terms modulo an integration by parts
\begin{equation}\label{21}
 \Gamma[H]=\frac{1}{2}\int_{ST_{2}}d^{4}z
               \left [ 4(\partial\zeta)H\;+\;2\zeta \partial H \right ].
\end{equation}

Just as in the bosonic case, we replace the partial derivative $\partial H$
in (\ref{21}) by the covariant one, i.e.,$(\partial-2D\zeta+\zeta D)H$,
according to (\ref{csd1'}), and the derivative of the superaffine
 connection $\partial\zeta$ by the superprojective connection
\begin{equation}\label{spc}
\Upsilon = -\partial\zeta - \zeta D\zeta.
\end{equation}

Now the integrand density in (\ref{21}) reads

\begin{equation}\label{22}
A_{ST}=4\left [ (-\Upsilon )H+
\frac{1}{2}\zeta\hat{\nabla}^{2}_{\zeta}H \right ].
\end{equation}

Here again the first term becomes globally defined when a holomorphic
superprojective connection $R$, with $sR=0,$ is introduced, since
the resulting expression is the difference of a generic superprojective
connection $R$ and a particular one, $\Upsilon$.\\
As mentioned previously, this is a superquadratic differential.
Indeed, using the transformation laws of $H, R, \zeta, \partial$
and $D$ given in \cite{DelGie',Gie} together with $(D_{\theta}w)^{2}=0$,
we get
\begin{eqnarray*}
(R\;+\;\partial\zeta\;+\;\zeta D\zeta)(\tilde{z},\tilde{\theta})
\;=\;\exp{(3w)}(R\;+\;\partial\zeta\;+\;\zeta D\zeta)(z,\theta).
\end{eqnarray*}

Thus $(R\;+\;\partial\zeta\;+\;\zeta D\zeta)H$
is the coefficient of a $(\frac{1}{2},\frac{1}{2})-$superdifferential.\\
This result expresses nothing more than the fact that super Beltrami
differentials and superquadratic differentials belong to dual spaces,
namely the tangent and the
cotangent spaces to the super Teichmller space $ST_{g}$ (see \cite{Nel}.)\\
To deal with the second term in $A_{ST}$, we have only to replace the
superaffine connection
$\zeta$ by the superabelian differential $(\zeta-\zeta_{0})$ where
$\zeta_{0}$ is a holomorphic
superaffine connection in the $0-$structure since $\hat
{\nabla}^{2}_{\zeta}H$ is already globally defined
as the covariant form of $\partial H$.\\
Considering all these changes together, we find the integrand of the
globally defined Polyakov action on the supertorus

\begin{equation}\label{24}
A_{ST}=4\left [ (R-\Upsilon )H+
          \frac{1}{2}(\zeta-\zeta_{0})\hat{\nabla}^{2}_{\zeta}H \right]
\end{equation}

Now, let us show that the expression above indeed solves the Ward identity
(\ref{1})
with the anomaly given in eq.(\ref{20}).\\
Using the transformation law in (\ref{19}) and
$sR=0, s\zeta_{0}=0$, we obtain

\begin{equation}\label{25}
s\Gamma[H,R]=\frac{1}{2}\int_{ST_{2}} d^{4}z
\left\{
{\cal A}(C,H,R)+D\phi-\bar{D}B_{ST}
\right\},
\end{equation}

where $\phi$ and $B_{ST}$ are

\begin{equation}\label{26}
\phi\;=\;T_{1}+T_{2}+T_{3}+T_{4}
\end{equation}
with
\begin{eqnarray}\label{27}
T_{1}&=&-(R-\Upsilon )(CDH\;+\;HDC) \nonumber \\
T_{2}&=&(\hat{\nabla}^{2}_{\zeta})(C\partial H\;-\;H\partial C) \nonumber \\
T_{3}&=&(\zeta D\zeta\;-\;\partial\zeta\;-\;\zeta\partial)(CDH\;+\;HDC)
\nonumber\\
T_{4}&=&(\hat{\nabla}^{2}_{\zeta})(DCDH)
\end{eqnarray}
and

 \begin{equation}\label{28}
  B_{ST}=4\left [ (R-\Upsilon )C+\frac{1}{2}
  (\zeta-\zeta_{0})
          \hat{\nabla}^{2}_{\zeta}C \right ]
 \end{equation}

Let us discuss the global definition of $\phi$
and $B_{ST}$ and then the behaviour of the corresponding integrals.
For this purpose we note that $B_{ST}$ is identical to $A_{ST}$ in (\ref{24})
up to the substitution of $H$ by $C$. Since both $H$ and $C$ transform
homogeneously under a change of coordinates, $B_{ST}$ shares with $A_{ST}$
the property of global definition.\\
Next we perform a coordinate transformation of $\phi$ according to the rules
in refs.\cite{DelGie',Gie} to get

\begin{eqnarray*}
\tilde{T}_{1}&=&\;\exp{(\bar{w})}T_{1} \\
\tilde{T}_{2}&=&\;
{\displaystyle
\exp{(\bar{w})}T_{2}+\exp{(\bar{w})}\left\{ \hat{\nabla}^{2}_{\zeta}
\left [Dw(CDH\;+\;HDC)\right ] \right \} }\\
\tilde{T}_{3}&=&\;
{\displaystyle
\exp{(\bar{w})}T_{3}+\exp{(\bar{w})}\left\{ \hat{\nabla}^{2}_{\zeta}
\left [Dw(CDH\;+\;HDC)\right ] \right \} } \\
\tilde{T}_{4}&=&\;
{\displaystyle
\exp{(\bar{w})}T_{4}+\exp{(\bar{w})}\left\{-2\hat{\nabla}^{2}_{\zeta}
\left [Dw(CDH\;+\;HDC)\right ] \right \} }.
\end{eqnarray*}

Finally, putting all terms $T_{i}, i=1,2,3,4$  together we find that $\phi$
is
globally defined since
\begin{eqnarray*}
\tilde{\phi}\;=\;\exp{(\bar{w})}\phi,
\end{eqnarray*}
i.e. $\phi$ is a $(0,\frac{1}{2})-$superdifferential.
Accordingly, each term in the r.h.s. of eq.(\ref{25}) is separately globally
defined. Moreover,
$\phi$ and $B$ are free from singularities, for they involve only non-singular
and single-valued fields (holomorphic differentials) on the supertorus $(g=1)$
according to super Riemann-Roch theorem (see \cite{Nel,RSV}).\\
Consequently, the last two integrals in $s\Gamma$ vanish thus leaving
only the anomaly. This finally
completes the proof that the Polyakov action proposed here with the integrand
in eq.(\ref{24}) indeed solves
the Ward identity (\ref{1}) with the anomaly (\ref{20}).

\section{Projection onto component fields}
The WZP action with the integrand density given in eq.(\ref{24}) is a
superspace integral involving superfields;
the corresponding component field expression is obtained in terms of
power series expansions of the superfields in the Grassman variables
$\theta$ and $\bar{\theta}$. In fact, the
holomorphic superfield
$R$ admits a $\theta-$expansion of the form \cite{Gie}

\begin{equation}\label{29}
R=\frac{i}{2}\chi+\theta \frac{1}{2}[r],
\end{equation}
where $\chi$ and $r$ (the projective connection introduced in sect.3) depend
on
the holomorphic variables $z$ and not on $\bar{z}$.
Moreover, in the WZ-supergauge we have \cite{DelGie'}

\begin{equation}\label{30}
H=\bar{\theta}\mu+\theta\bar{\theta}[-i\alpha] \hspace{1,5cm}; \hspace{1.5cm}
C=c+\theta[i\varepsilon].
\end{equation}

The space-time fields $\mu$ and $\alpha$ are the Beltrami coefficient and
its fermionic partner,
while $c$ and $\varepsilon$ denote the ghosts of ordinary diffeomorphisms and
local supersymmetry transformations respectively. However, the component field
expansion of the
affine connection $\zeta$ has not yet been given. An explicit solution to
the condition (\ref{4})
in the WZ-supergauge can be found in terms of ordinary space-time fields

\begin{equation}\label{31}
\hat{Z}=Z+\theta [i\partial Z \Psi]
\end{equation}

\begin{equation}\label{32}
\hat{\Theta}=\sqrt{\partial Z}[i\Psi+\theta(1-\frac{1}{2}\Psi\partial\Psi)]
\end{equation}

where $\Psi(z,\bar{z})$ is an anticommuting analytic field
and $Z$ is the projective coordinate
associated to the conformal structure of the underlying Riemann surface
$\Sigma$. Since $\partial Z$ is a 1-differential and $\theta$ is basically
viewed as a coordinate of the fiber of the spin bundle, i.e. transforming
as $\tilde{\theta} = \theta\sqrt{\partial\tilde{z}}$, the consistency of
eqs.(\ref{32}) implies that $\Psi$ transforms as a conformal field of
weight $-\frac{1}{2}$, i.e.
\begin{equation}\label{pstil}
\tilde{\Psi}(\tilde{z}) = \sqrt{\partial\tilde{z}}\Psi (z).
\end{equation}
{}From the defining relation (\ref{10}) and the expansion of $\hat{\Theta}$ in
(\ref{32}) we get
\
\begin{equation}\label{33}
\zeta=-i\frac{1}{2}\zeta^{0}-\theta[\frac{1}{2}\zeta^{1}]
\end{equation}

with
\begin{eqnarray}\label{34}
\zeta^{0}&=&\xi\Psi+2\partial\Psi = 2\nabla\Psi\nonumber \\
\zeta^{1}&=&\xi-\Psi\partial^{2}\Psi
\end{eqnarray}
where $\xi$ is the affine connection defined in the sect.3 and
 $\nabla$ the covariant derivative associated to it (see eq.(\ref{cd1})).\\
   Now it is easy to verify that $\zeta^{1}$ is an affine connection and that
$\zeta^{0}$ transforms homogeneously under a conformal
change of coordinates.
This result follows either by doing the explicit calculation through
the expressions (\ref{34}) or by projecting in components eq.(\ref{8}).
The component expansion of $\zeta_{0}$ is as follows
\footnote{Note that we have neglected the $\bar{\theta}$ and
$\theta\bar{\theta}$ components in the above expansions of $\hat{Z}$,
$\hat{\Theta}$, $\zeta$ and $\zeta^{0}$, whereas the superaffine connection
$\stackrel{\circ}{\gamma}_{\theta}$
considered in ref.\cite{AGN} contains this sort of contribution implied by
its factorized form.}
\begin{eqnarray}\label{35}
\zeta_{0}=-i\frac{1}{2}\zeta_{0}^{0}-\theta[\frac{1}{2}\zeta_{0}^{1}].
\end{eqnarray}

        The expansions (\ref{31}) and (\ref{32}) can be used to express the
Beltrami coefficient $\mu$ and the beltramino $\alpha$ through the expansion
(\ref{30}) in function of $Z$ and $\Psi$. The resulting expressions
\begin{eqnarray*}
\mu=\frac{\bar{\partial}Z}{\partial Z}(1+\Psi\partial\Psi)
-\Psi\bar{\partial}\Psi
\end{eqnarray*}
and
\begin{eqnarray*}
\alpha=2\bar{\partial}\Psi+\Psi\bar{\partial}\Psi\partial\Psi
-2\frac{\bar{\partial}Z}{\partial Z}\partial\Psi
+\frac{\bar{\partial}\partial Z}{\partial Z}\Psi
+\frac{\bar{\partial}Z}{\partial Z}\frac{\partial^{2}Z}{\partial Z}\Psi
\end{eqnarray*}
can be written as
\begin{equation}\label{36}
\bar{\partial}Z=\mu\partial Z+\frac{1}{2}\partial Z\Psi\alpha
\end{equation}
and
\begin{equation}\label{37}
\bar{\partial}\Psi=\frac{1}{2}\alpha(1-\Psi\partial\Psi)+\mu\partial\Psi
-\frac{1}{2}\partial\mu\Psi,
\end{equation}
 which are the supersymmetric extension of the well-known Beltrami
 equation.\\
By substituting the component field expressions (\ref{29}), (\ref{30}),
 (\ref{33}) and (\ref{35}) in the integrand density (\ref{24}) one finds
\begin{equation}\label{38}
\begin{array}{lcl}
& &\Gamma[\mu,\alpha;\zeta^{0},\zeta^{1}] =
{\displaystyle \int_{T_{2}} d^{2}z [ }

\chi\alpha+r\mu-(\partial\zeta^{1}-\frac{1}{2}(\zeta^{1})^{2}
-\frac{1}{2}\zeta^{0}\partial\zeta^{0})\mu-(\partial\zeta^{0}-\frac{1}{2}
\zeta^{0}\zeta^{1})\alpha \nonumber \\
& &-\frac{1}{2}(\zeta^{0}-\zeta_{0}^{0})(\partial\alpha+\frac{1}{2}
\zeta^{1}\alpha+
\partial\zeta^{0}\mu+\frac{1}{2}\zeta^{0}\partial\mu)-\frac{1}{2}
       (\zeta^{1}-\zeta_{0}^{1})
        (\partial\mu+\zeta^{1}\mu+\frac{1}{2}\zeta^{0}\alpha)

 {\displaystyle ]}.
 \end{array}
 \end{equation}

For $\Psi = 0$, we obtain the bosonic action given by eq.(\ref{16}).
The BRST transformations of the basic fields
follow from the superspace transformations, eqs.(\ref{19}),

\begin{eqnarray}\label{39}
s\mu&=&(\bar{\partial}-\mu\partial+\partial\mu)c+\frac{1}{2}\alpha
\varepsilon \nonumber \\
s\alpha&=&(\bar{\partial}-\mu\partial+\frac{1}{2}\partial\mu)\varepsilon+
c\partial\alpha
-\frac{1}{2}\alpha\partial c \nonumber \\
s\zeta^{0}&=& c\partial\zeta^{0}+\frac{1}{2}\partial c\zeta^{0}
+\partial\varepsilon+\frac{1}{2}\varepsilon\zeta^{1}
\nonumber \\
s\zeta^{1}&=& \partial(\partial c+c\zeta^{1}
-\frac{1}{2}\varepsilon\zeta^{0})
\end{eqnarray}

 and are completed by
\begin{eqnarray}\label{40}
s\chi=sr=s\zeta_{0}^{0}=s\zeta_{0}^{1}=0.
\end{eqnarray}

    From the superholomorphy equation (\ref{jed}) with $j=1$, the definition
 (\ref{10}) and the expansion (\ref{33}) it is straightforward to find the
 holomorphy conditions satisfied by $\zeta^{0}$ and $\zeta^{1}$
\begin{eqnarray}\label{41}
\bar{\partial}\zeta^{0}&=& \mu\partial\zeta^{0}+\frac{1}{2}\partial\mu
\zeta^{0}+\partial\alpha+\frac{1}{2}\alpha\zeta^{1} \nonumber \\
\bar{\partial}\zeta^{1}&=& \partial(\partial\mu+\mu\zeta^{1}
-\frac{1}{2}\alpha\zeta^{0}).
\end{eqnarray}
    With the help of eqs.(\ref{39},\ref{40}) and using properties (\ref{41})
( we recall that $\zeta_{0}^{0}$ and $\zeta_{0}^{1}$ are holomorphic fields ),
it is possible to verify that the action given by eq.(\ref{38}) is a solution
to the conformal Ward identity

\begin{eqnarray}\label{42}
s\Gamma[\mu,\alpha]= \frac{1}{2}\int_{T_{2}} d^{2}z
[{\cal A}+\bar{\partial}{\cal A}_{1}+\partial {\cal A}_{2}]
\end{eqnarray}
where ${\cal A}$ is the expression of the superdiffeomorphism anomaly in
component fields
(see eq.(3.78) in ref.\cite{Gie}, with the obvious changes of notations:
 ($ \rho \rightarrow \chi$ ; R$ \rightarrow $ r ) ),

 \begin{eqnarray}\label{43}
{\cal A}&=& 2r(c\partial\mu-\mu\partial c)-r\varepsilon\alpha +
(c\partial^{3}\mu -\mu\partial^{3}c)+\frac{3}{2}\chi (\varepsilon\partial\mu-
\mu\partial\varepsilon )+\frac{1}{2}\partial\chi\mu\varepsilon -
(\varepsilon\partial^{2}\alpha-\alpha\partial^{2}\varepsilon ) \nonumber \\
&-& \frac{3}{2}\chi(c\partial\alpha-\alpha\partial c)-\frac{1}{2}
\partial\chi c\alpha
\end{eqnarray}

and where ${\cal A}_{1}$ and ${\cal A}_{2}$ are
\begin{eqnarray}\label{44}
{\cal A}_{1}&=& 2(\chi\varepsilon + rc-\partial\zeta^{1}c-\frac{1}{2}\zeta^{1}
\partial c - \partial\zeta^{0}\varepsilon - \frac{1}{2}\zeta^{0}
\partial\varepsilon )+\zeta_{0}^{1}(\partial c+\zeta^{1}c+\frac{1}{2}
\zeta^{0}\varepsilon )\nonumber \\
&+& \zeta_{0}^{0}(\partial\varepsilon +
\frac{1}{2}\zeta^{1}\varepsilon + \partial\zeta^{0} c+\frac{1}{2}
\zeta^{0}\partial c) \nonumber \\
{\cal A}_{2}&=& \partial^{2}\mu c-\partial^{2}c\mu-\frac{1}{2}\chi\varepsilon
\mu+\frac{1}{2}\chi c\alpha - \partial (\varepsilon\alpha )
+\zeta^{1}(\partial\mu c-\mu\partial c-\varepsilon\alpha )\nonumber \\
&+& \partial\zeta^{0}(\varepsilon\mu -c\alpha)
+\frac{1}{2}\zeta^{0}(\partial\mu\varepsilon -\alpha\partial c).
\end{eqnarray}
    Now it remains to show that the last two integrals in the r.h.s.
of eq.(\ref{42}) vanish. Since they involves only non-singular
and single-valued fields on the torus it is sufficient to prove that
the expressions ${\cal A}_{1}$ and ${\cal A}_{2}$ are separately
globally defined
. First we note that ${\cal A}_{1}$ is identical to the integrand of the
expression (\ref{38}) when the Beltrami coefficients $\mu$ and $\alpha$
are replaced by the ghosts $c$ and $\varepsilon$ respectively. Hence, the
expression ${\cal A}_{1}$ is globally defined since, as $\mu$ and $\alpha$,
the fields $c$ and $\varepsilon$  transform homogeneously, i.e.
$\tilde{\varepsilon} =  \sqrt{\partial\tilde{z}}\varepsilon ;
\tilde{\alpha} = (\bar{\partial}\tilde{\bar{z}})^{-1}\sqrt{\partial
\tilde{z}}\alpha$.\\
The proof that ${\cal A}_{2}$ is also globally defined is achieved by
checking explicitly this property: $\tilde{\cal A}_{2} =
(\bar{\partial}\tilde{\bar{z}})^{-1}{\cal A}_{2}$, through the
fact that $\chi$ transforms
as a conformal field of weigth $\frac{3}{2}$ under a conformal change of
coordinates.

\section{Conclusion}

As an obvious result, the holomorphic factorization property of the Polyakov
action remains true on the supertorus.
Moreover, the results presented here illustrate the usefulness of the
covariant derivative,
making it straightforward to guess the global extension of an expression
initially given locally
 and constitute the first step towards the systematic study of the
Polyakov action on an arbitrary SRS. This generalization is highly
non-trivial since it requires
a thorough study of some properties (singularities and multi-valuedness) of
the fields it
would involve on the SRS. In particular the notion of the polydromy of
superdifferentials needs deeper understanding.
Relevant other geometrical notions necessary to this study will bring
interesting
insights into superconformal geometry on the SRS.
\newpage
\vskip 2.5truecm
{\bf \large{Acknowledgements}}  \\ [1cm]
It is a pleasure to thank A.Sebbar for sharing his knowledge of Riemann
surfaces and many fruitful discussions.
We are indebted to R.Zucchini for useful correspondence about his article.
We are also grateful to F.Gieres for his criticisms and
comments on a preliminary version of this work and to J.T.Donohue for a
careful reading of the manuscript.

\end{document}